\documentstyle[12pt]{article}

\begin{document}

\begin{center}
\indent
\vskip0.8cm

{\large \bf Perturbative analysis of generalized Einstein's theories}

\vspace{0.4cm}

\vspace{1cm}

J\'ulio C\'esar Fabris\footnote{e-mail: fabris@cce.ufes.br},\\
\vspace{0.3cm}
{\it Departamento de F\'{\i}sica, Universidade Federal do
Esp\'{\i}rito Santo,} \\
\vspace{0.1cm}
{\it Vit\'oria CEP 29060-900, Esp\'{\i}rito Santo, Brazil} \\
\vspace{0.4cm}
Richard Kerner\footnote{e-mail: rk@ccr.jussieu.fr},\\
\vspace{0.3cm}
{\it Laboratoire de Gravitation et Cosmologie Relativistes,} \\
\vspace{0.1cm}
{\it Universit\'e Pierre et Marie Curie, 75252 Paris Cedex 05, France} \\
\vspace{0.4cm}
and Jo\"el Tossa\footnote{e-mail: jtoss@syfed.bj.refer.org}\\
\vspace{0.3cm}
{\it Institut de Math\'ematiques et Sciences Physiques, Bo\^{\i}te Postale
613} \\
\vspace{0.1cm}
{\it Porto Novo, B\'enin} \\
\vspace{0.5cm}
\begin{abstract}
The hypothesis that the energy-momentum tensor of ordinary matter is not 
conserved separately, leads to a non-adiabatic expansion and, in many
cases, 
to an Universe older than usual. This may provide a solution for the
entropy 
and age problems of the Standard Cosmological Model. We consider two
different 
theories of this type, and we perform a perturbative analysis, leading to 
analytical expressions for the evolution of gravitational waves,
rotational 
modes and density perturbations. One of these theories exhibits
satisfactory 
properties at this level, while the other one should be discarded.
\end{abstract}
\end{center}

\vskip 0.2cm
\indent
\hskip 0.4cm

\section{Introduction}

The Cosmological Standard Model is based on the General Relativity theory, 
coupled to a perfect fluid energy-momentum tensor, with geometry
determined 
by the Robertson-Walker metric. It predicts an initial hot phase, when the 
baryogenesis and primordial nucleosynthesis take place, followed by the
matter 
dominated phase, when galactical structures are formed. This model can be 
considered as successful from many points of view, but it requires very 
stringent initial conditions. For example, the entropy observed today is
of 
the order of $10^{87} $ \cite{olive}, which can be also expressed by
saying 
that the actual density is very close to the critical density necessary
for 
the existence of a spatially flat Universe.
\newline
\indent
The thermodynamic equilibrium must also be imposed as an initial
condition, 
at least while the nucleosynthesis takes place, due to the existence of
the 
horizon distance smaller than the Hubble radius at that period. Moreover, 
the formation of structures are well explained if we admit the mean
density 
fluctuations of the order $\Delta \propto 10^{-5}$ when the radiation 
decouples from matter; but there is no clear mechanism explaining the 
amplification up to this value of small initial perturbations, originating
in
the very beginning of the Universe.
\par
The inflationary model, based on the existence of a de Sitter phase prior 
to the nucleosynthesis, copes with the entropy and horizon problem. It
also 
proposes a mechanism of generation and evolution of inhomogeneties. In
what
concerns the explanation of structure formation, inflation is in
competition
with other mechanisms, like those proposed by topological defect models.
In contrast to the latter, it proposes gaussian perturbations of quantum 
mechanical origin. Until very recently, the observational results of the 
COBE satellite concerning the anisotropies of the Cosmic Microwave
Background, 
seemed to favorize strongly the inflationary mechanism, but the gaussian 
nature of the perturbations has been questioned short time ago
\cite{Maqueijo}.
If this result is confirmed, it will become a serious drawback for the
inflationary models.
\par
Various other cosmological models try to propose alternative solutions to
the 
problems that remain open in the Standard Cosmological Model, without
using 
the inflationary paradigm. Many of them introduce ``exotic'' types of
matter
sources, or modify the dynamical equations\cite{thibault}. The models
introducing a ``cosmic string fluid'' \cite{davies} can 
lead to good results concerning the horizon problem, but exhibit
instabilities
at perturbative level \cite{sergio}. Another interesting possibility
consists 
in considering a ``non conservative'' theory of gravity, meaning a gravity 
theory in which the energy-momentum tensor does not conserve in the usual
way. 
The propositions of this type go back to the work of Rastall
\cite{rastall}, 
and have again attracted some attention recently \cite{Taha1,Taha2}. Such
a 
scheme leads to an older Universe in comparison with the standard results, 
creating at the same time more entropy during its evolution.
\par
On a purely formal level, these ``non-conservative'' theories are
equivalent 
to the introduction of exotic matter, whose main feature is negative
pressure.
More fundamentally, they lead to a radical modification of the dynamics of 
the Universe. We find frequently in the litterature statements explaining 
that such theories are incompatible with the equivalence principle. But
the 
equivalence principle means that particles follow geodesic curves
satisfying 
the equation $u^\mu u_{\nu;\mu} = 0$, and this requirement is totally 
independent of ${T^{\mu\nu}}_{;\mu} = 0$, and there is no incompatibility 
between the assumption ${T^{\mu\nu}}_{;\mu} \neq 0$ and the equivalence 
principle \cite{rastall}.
\par
Here we propose to study these theories from the perturbative point of
view,
paying special attention to the evolution of gravitational waves,
rotational
modes and density perturbations. The perturbative analysis intends to give 
more information concerning the structure formation in the realm of these 
theories, checking their stability at the same time. 
\par
In order to do this, we shall consider two different models. The first
one, 
referred to as ``Theory I'', is based on the Brans-Dicke theory, rewritten
in 
the Einstein frame. In this frame, matter is necessarily coupled to the
scalar 
field. Certain considerations on the physical aspects of the two different 
frames have been presented in reference \cite{frame}. In spite of serious 
objections against the theories of this kind, we will consider it here
mainly 
as a reference model. The second model, referred to as ``Theory II'', is
based 
on the original proposition due to Rastall, and in our opinion has better 
physical justification. The results obtained in the framework of this
model 
indicate that the improvements with respect to the standard scenario are
not compromised, at least at the perturbative level. This theory must be
also 
implemented with a convenient mechanism for generation and proper
evolution 
of inhomogeneties in an expanding globally homogenous and isotropic
Universe.
\par
In the next section, we present briefly the two theories and discuss their 
cosmological solution and some of its implications. In section 3, we
perform 
a perturbative study, considering perturbations of spin 2, 1 and 0. 
The comments and conclusions are contained in section 4.

\section{Generalized Einstein's theories}

The main point to be assured in order to solve the entropy problem without 
inflation, is to introduce a non adiabatic evolution of the Universe. This 
can be obtained by giving up the condition that the energy-momentum tensor 
has a vanishing covariant divergence. We can achieve this goal essentially
in 
two different ways. First, we can start with a non-minimal coupling
between 
the gravity and the scalar field on the one hand, and the matter on the
other 
hand (like e.g. in the Brans-Dicke theory); then, through a conformal 
transformation we can recover the minimal coupling between the gravity and
the 
scalar field, but with a more complicate coupling between matter, the
metric 
field and the scalar field. The second possible choice is to modify
Einstein's 
equations in such a way that the Bianchi identities are no longer
satisfied 
in their normal form, with a conserved energy-momentum tensor on the right
hand side, and in order to rectify the situation, we must modify the 
conservation laws for matter, which amounts to include new relationship
between the gravitational and other forms of energy.
\par
We will call the first choice {\it Theory I} and the second one 
{\it Theory II}. We shall consider now separately each of these
possibilities, 
analyzing their implications on the corresponding cosmological scenarios.

\subsection{Theory I}

We start with the Brans-Dicke theory coupled to the matter in the ordinary 
way
\begin{equation}
{\it L} = \sqrt{-g}\biggr(\phi R - 
\omega\frac{\phi_{;\rho}\phi^{;\rho}}{\phi}\biggl) 
+ {\it L_m}(g^{\mu\nu}) \quad .
\end{equation}
Now, we redefine the field $\phi$ and perform a conformal transformation
as in \cite{frame1}
\begin{eqnarray}
\label{ct1}
\phi &=&\frac{1}{2\kappa}e^{-2\beta\sigma} \quad ,\\
\label{ct2}
\tilde g_{\mu\nu} &=& e^{2\beta\sigma}g_{\mu\nu} \quad ,\\
\label{ct3}
\beta &=& \frac{1}{2\sqrt{\omega + \frac{3}{2}}} \quad .
\end{eqnarray}
Combining the equations (\ref{ct1}, \ref{ct2}, \ref{ct3}), we obtain a new 
lagrangian which can be written as :
\begin{equation}
{\it L} = \frac{\sqrt{-g}}{\kappa}\biggl(R - \sigma_{;\rho}\sigma^{;\rho}
\biggl) + {\it L_m}(e^{2\beta\sigma}g_{\mu\nu}) \quad ,
\end{equation}
where we have suppressed the tildes in order to simplify the notation. The
variational principle leads then to the following equations :
\begin{eqnarray}
R_{\mu\nu} - \frac{1}{2}g_{\mu\nu}R &=& \kappa \, T_{\mu\nu}
+ \sigma_{;\mu} \sigma_{;\nu} - \frac{1}{2}g_{\mu\nu} \sigma_{;\rho}
\sigma^{;\rho} \quad , \\
\Box\sigma &=& - \kappa \, \beta \, T \quad ,\\
{T^{\mu\nu}}_{;\mu} &=& \beta \, \sigma^{;\nu} \, T \quad ,
\end{eqnarray}
where $T = g^{\mu \nu} \, T_{\mu \nu}$ is the trace of the energy-momentum
tensor. Imposing the Robertson-Walker metric 
\begin{equation}
\label{m}
ds^2 = dt^2 - \frac{a^2(t)}{1 + \frac{k}{4}r^2}(dx^2 + dy^2 + dz^2) \quad
,
\end{equation}
and with the energy-momentum tensor of a perfect fluid, the above system 
reduces to the following three equations :
\begin{eqnarray}
3 \frac{{\dot a}^2}{a^2} + 3 \frac{k}{a^2} &=& \kappa \rho +
\frac{1}{2}(\dot\sigma)^2 \quad , \\
\ddot\sigma + 3\frac{\dot a}{a}\dot\sigma &=& -\kappa\beta(\rho
-3p) \quad ,\\
\label{em3}
\dot\rho + 3\frac{\dot a}{a}(\rho + p) &=& \beta\dot\sigma(\rho - 3p)
\quad .
\end{eqnarray}
The solutions to these equations for the case $k = 0$ are easily obtained
:
\begin{itemize}
\item $\alpha = -1$:
\begin{equation}
a \propto t^r \quad , \quad r = \frac{1}{8\beta^2} \quad ,
\quad \sigma = - \frac{1}{2\beta}\ln t \quad ;
\end{equation}
\item $\alpha = 0$:
\begin{equation}
a \propto t^r \quad , \quad r = \frac{2}{3 + 2\beta^2} \quad ,
\quad \sigma = - \frac{4\beta}{3 + 2\beta^2}\ln t \quad ;
\end{equation}
\item $\alpha = \frac{1}{3}$:
\begin{equation}
a \propto t^r \quad , \quad r = \frac{1}{2} \quad ,
\quad \sigma = constant \quad .
\end{equation}
\end{itemize}
Note that when $-\frac{3}{2} < \omega < \infty$, $\infty > \beta^2 > 0$.
\vskip 0.2cm
\par
These solutions display the following characteristic features :
\begin{enumerate}
\item The solution for the radiative Universe is the same as in General
Relativity, since the trace of the energy-momentum tensor is zero;
\item The solution for $\alpha = - 1$ exhibits
an inflationary behaviour when $\omega > \frac{1}{2}$. The same situation 
occurs in the original Brans-Dicke theory \cite{BDI};
\item The solution for the material phase does not exhibit an inflationary 
behavior and the scale factor evolves always slower than in the standard 
model ($ 0 < r < \frac{2}{3}$). This implies that the age of the Universe
in 
this case is less than in the standard case since $t_0 = r H_0$, where
$H_0$ 
is Hubble's constant measured today.
\end{enumerate}
\par
Concerning the entropy production, using the equation (\ref{em3}), and the 
relation $S = \frac{(\rho + p)a^3}{T}$  \cite{olive}, where $S$ and $T$
are 
the total entropy and temperature of the Universe, we obtain the following 
solution for $S$,
\begin{equation}
S = S_0\exp{\beta\sigma} = S_0\exp(-\frac{\beta^2}{3 + 2\beta^2}\ln t)
\quad .
\end{equation}
Hence the entropy always decreases with time during the material phase,
while
it remains constant during the radiative phase.

\subsection{Theory II}

The modified field equations are written with a a new parameter $\lambda$
as
\begin{equation}
\label{eq1}
R_{\mu\nu} - \frac{\lambda}{2} g_{\mu\nu} R = \kappa \, T_{\mu\nu} \quad .
\end{equation}
The divergence of (\ref{eq1}) leads to
\begin{equation}
\label{eq2}
{T^{\mu\nu}}_{;\mu} = \frac{1 - \lambda}{ 2 \kappa} \, R^{;\nu} \quad .
\end{equation}
Obviously these equations coincide with the standard ones when $\lambda =
1$.
We can write this system of field equations in an equivalent form as
\begin{eqnarray}
\label{fe1}
R_{\mu\nu} - \frac{1}{2}g_{\mu\nu}R &=& \kappa \, \biggr( T_{\mu\nu}
- \frac{\gamma - 1}{2} g_{\mu\nu} T \biggl) \quad ,\\
\label{fe2}
{T^{\mu\nu}}_{;\mu} &=& \frac{\gamma - 1}{2}T^{;\nu} \quad ,
\end{eqnarray}
\centerline{$ {\rm where} \, \ \ \, \lambda = \frac{2 - \gamma}{3 -
2\gamma}, \, \ \  
{\rm and} \, \ \ \, \ \  \gamma = \frac{3 \lambda - 2}{2 \lambda - 1} \, .
$}
\vskip 0.2cm
One checks easily that the covariant divergence of the energy-momentum
tensor 
vanishes and the equations become standard when $\gamma = 1$ (i.e. with 
$\lambda = 1$). This form of the system suggests that the modification 
amounts in fact to a re-defining the energy-momentum tensor by inclusion
of 
part of gravitational effects directly, and does not affect the geometric
framework of the theory.
\newline
\indent
This type of modification is less radical than the departure from the
Einstein-Hilbert lagrangian consisting in replacement of the scalar
curvature 
$R$ by a function $f(R)$, as in \cite{Duruisse}; the common feature
remaining
in both cases is that the extra geometrical terms may be interpreted as a
modification of the energy-momentum tensor.
\par
It is generally argued that theories of this type can not be derived from
a 
variational principle. However, we can define a matter lagrangian leading
to 
the equations (\ref{fe1}, \ref{fe2}). For example the matter lagrangian of 
the form :
\begin{equation}
{\it L_m} = \sqrt{-g}\biggr(x(\rho + p)u^\mu u^\nu g_{\mu\nu} + y\rho +
zp\biggl)
\end{equation}
leads to Einstein's equations when $x = 1 $, $y = - 1$ and $z = - 3$, and
to 
the generalized Einstein's equations when we choose $x = 1$, $y = -
\gamma$ 
and $z = - 3\gamma$.
\par
We look for solutions with spatially flat Robertson-Walker metric
(\ref{m}), 
where $a(t)$ is the scale factor describing the evolution of the Universe. 
We consider the energy-momentum tensor of a perfect fluid,
\begin{equation}
T^{\mu\nu} = (\rho + p)u^\mu u^\nu - pg^{\mu\nu} \quad ,
\end{equation}
where $\rho$, $p$ and $u^\mu$ denote respectively, the energy density, the 
pressure and the four-velocity of the fluid. We assume a barotropic
equation 
of state $p = \alpha\rho$, where $\alpha $ shall take on the values 
$-1$, $\frac{1}{3}$ and $0$, corresponding to the vacuum energy density, 
radiation and matter dominated equation of state respectively,
representing 
the most important cases for the primordial cosmological models. 
\par
For these three equations of state we can easily determine simple
solutions, namely,
\begin{itemize}
\item $\alpha = -1$:
\begin{eqnarray}
a(t) &\propto& \exp{Ht}\quad , \quad H = \sqrt{\frac{k}{3}(3 -
2\gamma\rho)}\quad,\\
\rho(t) &\propto& constant \quad ;
\end{eqnarray}
\item $\alpha = \frac{1}{3}$:
\begin{equation}
a(t) \, \propto \, t^\frac{1}{2} \quad ,   \, \ \  \,
\rho(t) \, \propto \, t^{-2} \quad ;
\end{equation}
\item $\alpha = 0$:
\begin{equation}
a(t) \, \propto \, t^{1 - \frac{\gamma}{3}} \quad , \, \ \ \,
\rho(t) \, \propto \, t^{-2} \quad ;
\end{equation}
\end{itemize}
Among the above solutions only the case $\alpha = 0$ presents an
improvement 
with respect to the standard model. The age of the Universe for this case 
can be evaluated as being equal to
\begin{equation}
t_0 = (1 - \frac{\gamma}{3}){H_0}^{-1} \quad ,
\end{equation}
leading to a Universe older than in standard mocdel for $\gamma < 1$. On
the 
other hand, using the equation (\ref{fe2}), and the same relations between 
$\rho$, $p$, $S$ and $T$ as in the previous section, we can deduce the 
entropy growth in the matter dominated Universe according to the power law
\begin{equation}
S = S_0t^{1 - \gamma} \quad .
\end{equation}
Hence, the condition $\gamma < 1$ implies at once an older Universe and
an increase of the total entropy during the matter dominated phase.
Moreover, 
the Universe exhibits an inflationary regime in the matter dominated era 
if $\gamma < 0$. We can also note that due to the particular coupling
between 
matter and gravity, it is possible to get negative values of the critical 
mass density parameter $\Omega$ and at same time a closed or spatially
flat 
Universe\cite {Taha1}. This feature is quite similar to the one displayed
by a model which couples ordinary matter with a string induced fluid, when  
closed Universe can evolve dynamically as an open one \cite{davies}.
However, 
it turns out that such models are hydrodynamically unstable \cite{sergio}.

\section{Perturbative analysis}

The Theory I considered above does not lead, in principle, to satisfactory 
results relative to the age of the Universe and the entropy generation.
The 
Theory II, on the other hand, gives acceptable results if the parameter 
$\gamma$ is restricted to the values $0 < \gamma < 1$. A perturbative 
analysis of these theories may give new insights into the physical 
consequences they lead to. In spite of previously shown negative results 
from the Theory I, we analyze it also from the perturbative point of view, 
for the sake of completeness. Our aim in both cases is to check if these 
theories can describe adequately the primordial gravitational waves, 
vectorial modes and density perturbations. In other words, we would like
to 
see if they can lead to growing vectorial modes, associated with
rotations, 
which is another drawback of the standard model. Also, if they can provide
a  
mechanism for the amplification of density perturbations compatible with 
observations, since it is generally assumed that they are very small in
the
beginning of the Universe, and must be amplified during all its history in
order to explain the observed large scale structures, as well as the
observed 
distortions of the Cosmic Background Radiation.
\par
Technically, we use the standard procedure of Lifschitz-Khalatnikov 
\cite{Lifshitz}. We isert into the field equations the perturbed
quantities 
${\tilde g}_{\mu\nu} = g_{\mu\nu} + h_{\mu\nu}$, $\tilde\rho = \rho 
+ \delta\rho$ and  $\tilde p = p + \delta p$, where $g_{\mu\nu}$, $\rho$ 
and $p$ are the background solutions around which small inhomogeneties are
created. Using the invariance of field equations with respect to the 
infinitesimal coordinate transformations, we may fix coordinate conditions 
eliminating unphysical degrees of freedom. Here we shall use the 
{\it synchronous coordinate condition} $h_{\mu0} = 0$. It is well known
\cite{peebles}, that after that a residual coordinate freedom remains, 
leading to a non-physical degree of freedom in the density perturbations. 
In our final results, we will consider the physical modes only. The 
derivation of the perturbed field equations is standard. In addition to
the 
considerations exposed above, we note that the perturbed quantities
$h_{ij}$ 
can be split into a sum of tensorial, vectorial and scalar modes as
follows:
\begin{equation}
h_{ij} = {h^{TT}}_{ij} + {h^{T}}_{(i|j)} + {h^s_1}_{|i|j} +
{h^s_2}\gamma_{ij} \quad ,
\end{equation}
where the first term is the transverse tracelless component (spin 2), the 
second one represents the vectorial mode (spin 1) and the last two ones
are 
identified as scalar modes (spin 0). The derivatives are defined in the 
three-dimensional spatial section and $\gamma_{ij}$ is the induced metric
of 
this spatial section. Moreover, it can be shown that the spatial
components
of the perturbed functions obey the eigenvalue equation $\nabla^2 Q = -
k^2Q$, 
where $k^2 = {\vec k}^2$, and $\vec k$ is the wave vector of the
perturbation
which is supposed to propagate as a plane wave.
\newline
\indent
We shall consider now the perturbed equations and their solutions in each
of 
the aformentioned theories. In all the following formulae, $k^2 = {\vec
k}^2$.

\subsection{Theory I}

\begin{itemize}
\item Spin 2:
\begin{equation}
\ddot h_{ij} - \frac{\dot a}{a}\dot h_{ij} +
\biggr(\frac{k^2}{a^2} +
4 \, \frac{\ddot a}{a} - (1 - \alpha)[3 \, \frac{{\dot a}^2}{a^2}
- \frac{1}{2} \frac{{\dot\sigma}^2}{2}]\biggl)h_{ij} = 0 \quad ;
\end{equation}
\item Spin 1:
\begin{equation}
\dot{\delta u^i} + \biggr[(2 - 3\alpha) \, \frac{\dot a}{a}
+ \beta\dot\sigma \, (1 - 3\alpha) \biggl] \, \delta u^i = 0 \quad ;
\end{equation}
\item Spin 0:
\begin{eqnarray}
\label{ep1}
\ddot h + 2 \frac{\dot a}{a} \, \dot h 
- 4\dot\sigma \dot\chi - \biggr(3 \, (\frac{{\dot a}^2}{a^2})
- \frac{1}{2} \, {\dot\sigma}^2\biggl)(1 + 3 \alpha) \Delta &=& 0 \quad
;\\
\label{ep2}
\ddot\chi + 3\frac{\dot a}{a} \, \dot\chi - \nabla^2\frac{\chi}{a^2}
- \frac{1}{2}\dot\sigma\dot h
+ \biggr(3 \, (\frac{{\dot a}^2}{a^2})
- \frac{1}{2} \,{\dot\sigma}^2\biggl)(1 - 3\alpha) \, \Delta &=& 0\quad ;
\\
\label{ep3}
\dot\Delta + (1 + \alpha)\biggr(\Theta - \frac{1}{2}\, \dot h\biggl)
- \beta(1 - 3\alpha)\dot\chi &=& 0\quad ;\\
\label{ep4}
(1 + \alpha)\dot\Theta +
(1 + \alpha)\biggr((2 - 3\alpha)\frac{\dot a}{a} + \beta(1 - 3\alpha)
\dot\sigma\biggl)\Theta &+&\nonumber\\ 
+ \frac{\nabla^2}{a^2}\biggr(\alpha\Delta +
\beta(1 - 3\alpha) \, \chi\biggl) &=& 0 \quad .
\end{eqnarray}
\end{itemize}
In the expressions above, $\Delta = \frac{\delta\rho}{\rho}$,
$\Theta = \delta u^i_{,i}$, $h = \frac{h_{kk}}{a^2}$ and $\chi = 
\delta\sigma$.
\par
The equation describing the tensorial mode admits a simple solution when 
$\alpha = 0$ in terms of the conformal time defined as $dt = ad\eta$:
\begin{equation}
h^{TT} \propto \eta^{\frac{1 + r}{2(1 - r)}}c_{\pm}J_{\pm\nu}(k\eta)
\end{equation}
where $\nu = \frac{3 - 2\beta^2}{2(1 + 2\beta^2)}$, $J_{\pm \nu}$ are
Bessel 
functions with two possible values of index, $\pm \nu$, and $c_\pm$ are
constants of integration. The results are essentially similar to those 
obtained in the General Relativity framework for $\alpha = \frac{1}{3}$ 
\cite{grish}, and in the original Brans-Dicke theory \cite{plinio} for 
$\alpha = - 1$. The vectorial mode equation also has a simple solution
with
\begin{equation}
\delta u^i \propto t^{\frac{4(\beta^2 - 1)}{3 + 2\beta^2}}\quad .
\end{equation}
Growing rotational modes can appear if $\beta^2 >1$. This leads to the 
restriction $ -\frac{3}{2} < \omega < -\frac{5}{4}$, implying 
$\frac{1}{3} < r < \frac{2}{5}$. s It should be noted that such values of 
$r$ correspond to a very young Universe.
\par
For the density perturbations (spin 0), the solution is exactly the same
as 
for the radiative phase in General Relativity. As it is well known, there 
is no significant amplification of the perturbations during this phase 
 \cite{grish}. Nevertheless, unlike in General Relativity, we can have 
non-zero solutions for the perturbations during the de Sitter phase.
Choosing 
$\alpha = - 1$ in the perturbed equations, we find first that $\Delta = 4 
\beta \chi$. Using this relation, we can determine a third order
diferential 
equation satisfied by $\Delta$:
\begin{eqnarray}
\stackrel{...}{\Delta} + \biggr(5  \frac{\dot a}{a} - 
\frac{\ddot\sigma}{\dot\sigma}\biggl) \ddot\Delta +\nonumber\\
+ \biggr[(\frac{k^2}{a^2}) +
3\biggr(\frac{\ddot a}{a} +
(\frac{\dot a^2}{a^2}) - \frac{\dot a}{a}\frac{\ddot\sigma}{\dot\sigma}
\biggl) - 2\dot\sigma^2 +
16\beta^2\biggr(3(\frac{\dot a^2}{a^2}) -
\frac{1}{2}\dot\sigma^2\biggl)\dot\Delta + \nonumber \\
+ \biggr[-\frac{\ddot\sigma}{\dot\sigma}(\frac{k}{a})^2 +
16\beta^2\biggr(3\frac{\dot a}{a}(2\frac{\ddot a}{a} - \frac{\ddot\sigma}
{\dot\sigma}\frac{\dot a}{a}) -
\frac{\dot a}{a}\dot\sigma^2 - \frac{1}{2}\dot\sigma\ddot\sigma\biggl)
\nonumber \\
+ 4\beta\dot\sigma\biggr(3(\frac{\dot a}{a})^2 - \frac{1}{2}\dot\sigma^2
\biggl)\biggl]\Delta = 0
\end{eqnarray}
As usual, $\Delta \propto \frac{1}{t}$ is a solution that is related to
the 
residual coordinate freedom. Inserting $\Delta = \frac{1}{t}\zeta$ and
using 
the conformal time such that $dt = ad\eta$, then writing $\zeta =
\eta^p\xi$, 
where $p = \frac{3}{2}$, we find the following equation for $\xi$:
\begin{equation}
\xi'' + \frac{\xi'}{\eta} +
\biggr[1 - \frac{1}{4}(\frac{1 + 8\beta^2}{1 -
8\beta^2})^2\frac{1}{\eta^2}
\biggl]\xi = 0 \quad .
\end{equation}
This is a Bessel equation. Turning back to $\Delta$, we have the solution:
\begin{equation}
\Delta = \frac{1}{\eta^{s+1}}\int \eta^\frac{3}{2}c_\pm 
J_{\pm\nu}(k\eta)d\eta \quad ,
\end{equation}
with $\nu = \frac{1}{2}\frac{1 + 8\beta^2}{1 - \beta^2}$ and $s =
\frac{1}{8\beta^2 - 1}$.
\par
This solution is not valid for $n = 1$ when this is the case, we find an 
Euler equation,:
\begin{equation}
\stackrel{...}{\Delta} + 6\frac{\ddot\Delta}{t} +
(k^2 + 6)\frac{\dot\Delta}{t^2} + k^2\Delta = 0 \quad .
\end{equation}
Discarding the solution connected with the residual coordinate freedom 
we have :
\begin{equation}
\Delta = t^{- 1 \pm \sqrt{1 - k^2}} \quad .
\end{equation}
The main feature of these solution is that asymptotically, for small $k$ 
(large wavelength of the perturbations), it does not contain growing
modes.
\par
For $\alpha = 0$, we can find solutions in the long wavelength limit,
$k \rightarrow 0$. They read :
\begin{equation}
\Delta \propto t^{p} \quad, \quad p = \frac{1}{2(3 + 2\beta^2)}
\biggr(- 1 + 6\beta^2 \pm (5 - 2\beta^2)\biggl)
\quad .
\end{equation}
The maximum value of $p$ appears when $\omega \rightarrow \infty$, leading 
to $p = \frac{2}{3}$. In this limit the theory coincides with General 
Relativity. Hence, during the material phase, the perturbations grow less 
rapidly than in the standard model.
\subsection{Theory II}
For this theory, the perturbed equations read
\begin{itemize}
\item Spin 2:
\begin{equation}
\ddot h_{ij} - \frac{\dot a}{a}\dot h_{ij} +
\biggr((\frac{k^2}{a^2}) - 2\frac{\ddot a}{a}\biggl)h_{ij} = 0 \quad ;
\end{equation}
\item Spin 1:
\begin{equation}
(1 + \beta)\dot{\delta u^i} + (1 + \beta)(2 - 3\beta)\frac{\dot a}{a}
\delta u^i = 0 \quad ;
\end{equation}
\item Spin 0:
\begin{eqnarray}
\ddot h + 2\frac{\dot a}{a}\dot h &=& 6(\frac{\dot a^2}{a^2}) (1 + \beta)
\Delta \quad , \\
\dot\Delta + (1 + \beta)(\Theta - \frac{1}{2}\dot h) &=& 0 \quad , \\
(1 + \beta)\dot\Theta + (1 + \beta)(2 - 3\beta)\frac{\dot a}{a}\Theta &=&
\beta(\frac{k^2}{a^2}) \Delta \quad ,
\end{eqnarray}
\end{itemize}
where here $\beta = \frac{(5 - 3\gamma)\alpha + \gamma - 1}{3 - \gamma
- 3(1 - \gamma)\alpha}$. Hence, the equations (so, the solutions) are 
similar to those of the standard model, but with a ``modified equation
of state'' in which $\beta$ replaces $\alpha$. We can verify that, for 
$\alpha = -1$, $\beta = -1$, $\alpha = \frac{1}{3}$, $\beta =
\frac{1}{3}$, 
and $\alpha = 0$, $\beta = \frac{\gamma - 1}{3 - \gamma}$. The results
are the same as in the standard model for the false vacuum and radiative
phases. During these periods, the rotational modes decrease (or simply
does
not exist, as in for the false vacuum case), and the density perturbations
exist only in the radiative case, but are amplified very moderately.
\par
Let us analyze now the case $\alpha = 0$. For $0 < \gamma < 3$ (so that
the 
weak energy condition is satisfied), $ -\frac{1}{3} < \beta < \infty$.
For $0 < \gamma < 1$ (growing entropy and older Universe), we have
$- \frac{1}{3} < \beta < 0$ (the effective pressure is negative, but
the strong energy condition is not violated). We consider now the
solutions 
for the three different modes separately. The scale factor behaves as 
$a \propto t^{1 - \frac{\gamma}{3}}$, or, in terms of the conformal time, 
$a \propto \eta^{\frac{3 - \gamma}{\gamma}}$,
\begin{itemize}
\item Spin 2:
The solutions can be written as,
\begin{equation}
h_{ij} \propto \eta^{\frac{1}{2}\frac{6 - \gamma}{\gamma}}
J_\nu(k\eta) \quad , \quad \nu = \frac{6 - 3\gamma}{2\gamma} \quad .
\end{equation}
For large wavelengths, it behaves asymptotically as
\begin{equation}
h_{ij} \propto \eta^\frac{6 - 2\gamma}{\gamma} \quad .
\end{equation}
In the interval of interest, the gravitational waves are amplified, but
not in a exponential way.
\item There are growing rotational perturbations during the material era 
provided that
\begin{equation}
3\beta - 2 > 0 \Rightarrow \beta > \frac{2}{3} \Rightarrow \gamma >
\frac{9}{5} \quad .
\end{equation}
However, this leads to an young Universe and a decreasing entropy($\gamma
>
1$);
\item For the density perturbations, we can note that:
\begin{enumerate}
\item For $\gamma > 1$, the solutions are equivalent to those of the
standard 
model with an equation of state $p = \alpha\rho$, $\alpha > 0$, namely,
\begin{equation}
\label{perturb}
\Delta \propto \frac{1}{\eta^\frac{3}{\gamma}}
\int \eta^\frac{5}{2}c_\pm J_{\pm\nu}(k\eta) \quad ,
\quad \nu = \frac{6 - 3\gamma}{2\gamma} \quad .
\end{equation}
\item For $0 < \gamma < 1$ and $\gamma < 0$, the solution are the same as
in General Relativity,
with $- \frac{1}{3} < \alpha < 0$, which amounts to replacing the Bessel's
functions by the modified Bessel's functions  (see, e.g. \cite{fabris}) :
\begin{equation}
\Delta \propto \frac{1}{\eta^\frac{3}{\gamma}}
\int \eta^\frac{5}{2}\biggr(c_1 I_\nu(k\eta) +
c_2 K_\nu(k\eta)\biggl) \quad .
\end{equation}
\end{enumerate}
In both cases, the long wavelength limit yields the same asymptotic
behaviour
\begin{equation}
\Delta \propto \eta^2 \propto t^{\frac{2}{3}\gamma} \quad.
\end{equation}
We can see that the perturbations, in this limit, and with $\gamma > 0$, 
grow faster than in the Standard Model when the Universe is younger, and 
slower when the Universe is older. In both cases, we can expect the final 
amplification of the perturbations to be of the same order. If $\gamma <
0$, 
the density perturbations are decreasing.
\end{itemize}
\section{Conclusions}
Models where the momentum-energy tensor is not conserved {\it per se} may
give 
reasonable results while coping with some traditional problems posed by
the
Standard Cosmological Model. Here we have considered two kinds of such 
theories. The first one is based on a non-trivial interaction between the
matter and the scalar field, which can be obtained from the Brans-Dicke 
theory through conformal transformation. The second one is based on a
direct 
modification of Einstein's theory, via introduction of the parameter
$\gamma$ 
in Einstein's equations.
\par
The first theory is marred by problems from the Cosmological point of view 
even in the background solution, since it predicts an Universe younger
than 
the one predicted by the standard model, and decreasing entropy during the 
matter dominated phase. The second theory, in contrast, gives the opposite 
result when $0 < \gamma < 1$. For $\gamma < 0$ it predicts an inflationary 
behaviour during the matter-dominated phase.
\par
Turning to the structure formation problem, we can state the following
results.
\begin{enumerate}
\item In the realm of the first theory, the gravitational mode has
qualitatively the same behaviour than in the standard model. The
rotational
mode exhibit a growing solution but for a very young Universe.
But, contrary to the standard model, there exist density perturbations
in the false vacuum case, which, however, have a decreasing behaviour
in the long wavelength limit. In the radiative phase, the behaviour is
the same than in the standard model, but during the matter dominated
phase, the density perturbations grow less rapidly, which is even more 
serious drawback if we remember that the age of the Universe is also lower 
than in the standard case;
\item In the second theory, we have the same behaviour that within the 
standard model if $0 < \gamma < 1$ for gravitational and rotational modes. 
Rotational modes may exhibit a growing tendency for $\gamma > \frac{9}{5}$
but this leads to a very young Universe and the entropy decreasing during
the 
matter dominated phase. The behaviour of the density perturbations remains 
unchanged with respect to the standard results during the false vacuum and 
radiative phases. But during the matter dominated phase there is a fast
growing mode (in the long-wavelength limit) if $\gamma > 1$, and a slower 
one if $\gamma < 1$. The amplification is essentially the same in this
phase 
due to the modifications in the evaluation of the age of the Universe.
\end{enumerate}
The first theory does not lead to acceptable results neither for the 
background solution, nor at the perturbative level, whereas the second one 
gives interesting background solutions and, at least, is not disqualified 
by the perturbation analysis. This can be judged rather encouraging, since 
as in the standard model, the second theory needs a mechanism for
generation 
and amplification of perturbations in phases prior to the matter dominated 
era; but, on the other hand, we remember that certain models which give 
interesting background solutions are entirely compromised at the
perturbative 
level, as is the case of the models based on string like fluid
\cite{sergio}, 
and which is not the case here.
\newline
\indent
For $\gamma < 1$, this theory leads to a reasonable entropy production, to 
an older Universe and to growing modes for density perturbations in the 
material phase.
\par
Is it possible to extract some observational limit for the parameter
$\gamma$ 
from the above perturbative analysis? In principle, yes. We could employ
for
example the observations on the anisotropies of the cosmic microwave 
background, measured by the COBE satellite. However, in order to do that, 
we must have a complete scenario at our disposal, including the above 
mentioned mechanism for the origin and amplification of the perturbations. 
In our case, this should be done without the help of inflation in order to 
be coherent with the proposals of the model. 
\par
Even without that, we can extract some observational informations. One 
important observational parameter is the spectral index $n$,
characterizing
the spectrum of the perturbations. If $n = 1$ we have the so called
Harrison-Zeldovich spectrum: all perturbations, disregarding their
wavelength, have the same amplitude when they enter the horizon. The
COBE's 
results indicate a nearly Harrison-Zeldovich spectrum, with $n=1$
representing 
possible value.
\newline
\indent
The spectral index can be evaluated from the two point correlation
function. 
It gives the following result \cite{kaiser,brandenberg},
\begin{equation}
n = 1 + \frac{d\ln\delta^2}{d\ln k} \quad ,
\end{equation}
where $\delta^2$ is proportional to the two point correlation function,
\begin{equation}
\vert\Delta(\vec k,t)\vert \equiv k^3\int \frac{d^3x}{(2\pi)^3}
e^{i\vec k.\vec x}\langle\Delta(\vec x,t)\Delta(0,t)\rangle \quad.
\end{equation}
Using the solutions (\ref{perturb}), considering the long wavelength limit 
with the solution that leads to the Bunch-Davies vacuum (for details, see 
\cite{kaiser}), we have
\begin{equation}
\delta \propto k^{3 - 2\nu} \quad ,
\end{equation}
leading to
\begin{equation}
n = 1 + 6\frac{(\gamma - 1)}{\gamma} \quad .
\end{equation}
The observational limits of $n$ seem to indicate a value very close to 
$n = 1$. Taking for simplicity (since we are only making a crude
estimation) 
$0.8 < n < 1.2$,
we obtain $0.96 < \gamma < 1.03$. Following this simplified analysis, we 
find that the Universe today must be, from the Theory II, a little older 
than the one predicted by the standard model, with a very low production
of 
entropy during the matter dominated phase.

\end{document}